\documentclass[twocolumn,showpacs,prd]{revtex4}
\usepackage{amsmath}
\usepackage{mathrsfs,bm,multirow}
\usepackage{longtable,lscape}
\usepackage{txfonts}
\usepackage{amssymb}
\usepackage{indentfirst}
\usepackage{graphicx,,booktabs}
\usepackage{color}
\usepackage{amssymb}
\begin{document}

\title{Finding the $0^{--}$ Glueball}
\author{Cong-Feng Qiao$^{1,2}$}
\email{qiaocf@ucas.ac.cn}
\author{Liang Tang$^{1}$}
\email{tangl@ucas.ac.cn}
\affiliation{$^1$School of Physics, University of Chinese Academy of Sciences - YuQuan Road 19A, Beijing 100049, China\\
$^2$CAS Center for Excellence in Particle Physics, Beijing 100049, China}

\begin{abstract}

 With appropriate interpolating currents the mass spectrum of $0^{--}$ oddball is obtained in the framework of QCD Sum Rules (QCDSR). We find there are two stable oddballs with masses of $3.81 \pm 0.12 \, \text{GeV}$ and $4.33 \pm 0.13 \, \text{GeV}$, and analyze their possible production and decay modes in experiments. Noticing that these $0^{--}$ oddballs with unconventional quantum number are attainable in BESIII, BELLEII, PANDA, Super-B and LHCb experiments, we believe the long search elusive glueball could be measured shortly.

\end{abstract}
 \maketitle

Quantum Chromodynamics (QCD) is the underlying theory of hadronic interaction. In the high energy regime, it has been tested up to 1\% level due to the asymptotic freedom \cite{Gross:1973id}. However, the nonperturbative aspect related to the hadron spectrum is difficult to be calculated from the first principle because of the confinement \cite{Wilson:1974sk}. A unique attempt in understanding the nonperturbative aspect of QCD is to study the glueball ($gg$, $ggg$, $\cdots$), where the gauge field plays a more important dynamical role than in ordinary hadrons. This has intrigued much interest in theory and experiment for quite long time.

In the literature, many theoretical investigations on glueball were made through various techniques, including lattice QCD \cite{Wilson:1974sk,  Mathieu:2008me, Yang:2013xba}, flux tube model \cite{Isgur:1984bm}, MIT bag model \cite{Chodos:1974je, Jaffe:1975fd}, Coulomb Gauge model \cite{Szczepaniak:1995cw} and QCD Sum Rules (QCDSR) \cite{Shifman, Reinders:1984sr, twogluon0++, Huang:1998wj, Narison:1996fm, twogluon0-+, Latorre:1987wt, Lu:1996tp, Hao:2005hu}. Of these techniques, the model independent QCDSR, developed more than thirty years ago by Shifman, Vainshtein and Zakharov (SVZ) \cite{Shifman}, has some peculiar advantages in the study of hadron phenomenology. Its starting point in evaluating the properties of the ground-state hadron is to construct the current, which possesses the foremost information about the concerned hadron, like quantum numbers and constituent quark or gluon. By using of the current, one can then construct the two-point correlation function, which has two representations: the QCD representation and the phenomenological representation. Equating these two representations, the QCDSR will be formally established.

In the framework of QCDSR, the two-gluon glueballs with quantum numbers of $0^{++}$ \cite{twogluon0++, Huang:1998wj, Narison:1996fm}  and $0^{-+}$ \cite{Narison:1996fm, twogluon0-+} have been studied extensively in the literature. Note that these glueballs were also constructed and investigated through tri-gluons \cite{Latorre:1987wt, Lu:1996tp, Hao:2005hu}, which is enlightening for the research in this work.

Although glueball has been searched for many years in experiment, so far there has been no definite conclusion about it, mainly due to the following  three reasons: the mixing effect between glueballs and quark states, the lack of the glueball production mechanism, and the lack of the necessary knowledge about glueball decay properties. Of these difficulties, from the experimental point of view, the most outstanding obstacle is how to isolate glueball from the mixed quarkonium states ( $q \bar{q}$). Fortunately, there is a class of glueballs, the unconventional glueballs, which with quantum numbers unaccessible by quark-antiquark bound states can avoid such problem. The quantum numbers of those glueballs include $J^{PC} = 0^{--}$, $0^{+-}$, $1^{-+}$, $2^{+-}$, $3^{-+}$, and so on. Note, according to C-parity conservation, glueballs with negative C-parity cannot be reached  by two gluons, but have to be composed of at least three gluons. In the literature the term oddball has been used to describe 3 gluon glueballs having unconventional quantum numbers as well as 3 gluon glueballs with odd J, P, C having conventional quantum numbers. To unify and avoid confusion, we propose using the term oddball to simply refer to glueballs with 3 gluons.

Among oddballs, special attention ought be paid to the $0^{--}$ ones, since they are relatively light and their quantum number enables their production in the decays of vector quarkonium or quarkoniumlike states easier. The aim of this letter is to evaluate the mass spectrum of $0^{--}$ oddball and analyze the feasibility of finding it in experiment.

In order to calculate the mass spectrum of $0^{--}$ oddball, one has to construct the appropriate current for it. In practice a number of currents satisfy the unconventional quantum number. However, after imposing the constraints of gauge invariance, Lorentz invariance and $SU_c(3)$ symmetry, only a limited number of currents remain. They are
\begin{eqnarray}
j^A_{0^{--}}(x) & \!\!\!\! = \!\!\!\! & g_s^3 d^{a b c} [g^t_{\alpha \beta} \tilde{G}^a_{\mu \nu}(x)][\partial_\alpha \partial_\beta G^b_{\nu \rho}(x)][G^c_{\rho \mu}(x)]\, , \label{current-A} \\
j^B_{0^{--}}(x) & \!\!\!\! = \!\!\!\! & g_s^3 d^{a b c} [g^t_{\alpha \beta} G^a_{\mu \nu}(x)][\partial_\alpha \partial_\beta \tilde{G}^b_{\nu \rho}(x)][G^c_{\rho \mu}(x)]\, , \label{current-B} \\
j^C_{0^{--}}(x) & \!\!\!\! = \!\!\!\! & g_s^3 d^{a b c} [g^t_{\alpha \beta} G^a_{\mu \nu}(x)][\partial_\alpha \partial_\beta G^b_{\nu \rho}(x)][\tilde{G}^c_{\rho \mu}(x)]\, , \label{current-C} \\
j^D_{0^{--}}(x) & \!\!\!\! = \!\!\!\! & g_s^3 d^{a b c} [g^t_{\alpha \beta} \tilde{G}^a_{\mu \nu}(x)][\partial_\alpha \partial_\beta \tilde{G}^b_{\nu \rho}(x)][\tilde{G}^c_{\rho \mu}(x)]\, ,\label{current-D}
\end{eqnarray}
where $a$, $b$, $c$ are color indices, $\mu$, $\nu$, $\rho$, $\alpha$, $\beta$ are Lorentz indices, $d^{a b c}$ stands for the totally symmetric $SU_c(3)$ structure constant, $g^t_{\alpha \beta}= g_{\alpha \beta}- \partial_\alpha \partial_\beta/\partial^2$, $G^a_{\mu \nu}$ denotes the gluon field strength tensor, and $\tilde{G}^a_{\mu \nu}$ is the dual gluon field strength tensor defined as $\tilde{G}^a_{\mu \nu} = \frac{1}{2} \epsilon_{\mu \nu \kappa \tau} G^a_{\kappa \tau}$\ .  Hereafter, for simplicity the four $0^{--}$ currents in Eqs.(\ref{current-A})-(\ref{current-D}) will be referred as case A to D respectively, and they will be all taken into account in our analysis.

With the currents of (\ref{current-A})-(\ref{current-D}), the two-point correlation functions can be readily established, i.e.
\begin{eqnarray}
\Pi(q^2) = i \int d^4 x e^{i q \cdot x} \langle 0 | T \bigg\{ j_{0^{--}}(x), j_{0^{--}}(0)\bigg\}| 0 \rangle \, ,
\end{eqnarray}
where $|0 \rangle$ denotes the physical vacuum. The QCD side of the correlation function can be obtained through the Operator Product Expansion (OPE) and reads as
\begin{eqnarray}
\mathrm{\Pi^{QCD}}(Q^2) &\!\!\!\!=\!\!\!\!& a_0 Q^{12} \ln\frac{Q^2}{\mu^2} + b_0 Q^{8} \langle \alpha_s G^2 \rangle  \nonumber \\
&\!\!\!\!+\!\!\!\!& \left( c_0 + c_1 \ln\frac{Q^2}{\mu^2} \right) Q^{6} \langle g_s G^3 \rangle + d_0 Q^4 \langle \alpha_s G^2 \rangle^2 \ . \label{correlation-function-QCD}
\end{eqnarray}
Here, $\langle \alpha_s G^2 \rangle$, $\langle g_s G^3\rangle$, and $\langle \alpha_s G^2 \rangle^2$ represent two-gluon, three-gluon, and four-gluon condensates respectively; $\mu$ is the renormalization scale; and $Q^2 \equiv -q^2 > 0$. For simplicity, we use $a_0$, $b_0$, $c_0$, $c_1$ and $d_0$ to represent the Wilson coefficients of opertators with different dimensions in Eq.(\ref{correlation-function-QCD}). After a lengthy calculation, the Wilson coefficients are obtained as follows:
\begin{eqnarray}
\begin{aligned}
&a_0^i = \frac{487 \alpha_s^3}{143\times 2^6 \times 3^3 \pi}\ , \  b_0^i = -\frac{5 \pi}{36}\alpha_s^2\; ,
c_0^A =  - \frac{205}{12} \pi \alpha_s^3\ , \\
&c_1^A = - \frac{775}{144} \pi \alpha_s^3\ , \ c_0^B = - \frac{2065}{48} \pi \alpha_s^3\; ,  c_1^B =  - \frac{1075}{96} \pi \alpha_s^3\ , \\
&c_0^C = \frac{2275}{72} \pi \alpha_s^3\ , \ c_1^C = \frac{2125}{144} \pi \alpha_s^3\; , \ c_0^D = - \frac{1045}{144} \pi \alpha_s^3\ , \\
&c_1^D = - \frac{25}{32} \pi \alpha_s^3\; ,
\ d_0^j = 0 \ ,  \ d_0^D = - \frac{5}{9} \pi^3 \alpha_s\; ,
\end{aligned}
\end{eqnarray}
where the superscript $i$ runs from A to D and $j$ for A to C, with A, B, C and D corresponding to the four currents in Eqs.(\ref{current-A})-(\ref{current-D}), respectively. Notice that there are symmetries within Wilson coefficients $a_0^i$, $b_0^i$, and $d_0^j$. Since the position and the number of $\tilde{G}$ in Eqs.(\ref{current-A})-(\ref{current-D}) do not influence the perturbative and $\langle \alpha_s G^2 \rangle$ contributions,
$a_0^i$($b_0^i$) are identical for all cases; whereas they influence the $\langle g_s G^3 \rangle$ term, and hence $c_0^i$($c_1^i$) are different. Moreover, $\langle\alpha_s G^2 \rangle^2$ term involves no loop contribution, $d_0^j$ are governed by the number of $\tilde{G}$, so they are equal.

On the phenomenological side, adopting the pole plus continuum parameterization of the hadronic spectral density, the imaginary part of the correlation function can be saturated as:
\begin{eqnarray}
\frac{1}{\pi} \mathrm{Im\Pi^{phe}}(s) = f_G^2 M_{0^{--}}^{12} \delta(s - M_{0^{--}}^2) + \rho (s) \theta(s - s_0) \; .
\end{eqnarray}
Here $\rho(s)$ is the spectral function of excited states and continuum states above the continuum threshold $\sqrt{s_0}$, $M_{0^{--}}$ represents the mass of $0^{--}$ oddball, $f_G$ stands for the coupling parameter defined by the following matrix element,
\begin{eqnarray}
\langle 0 | j_{0^{--}}(0)| G\rangle = f_G M_{0^{--}}^6 \; .
\end{eqnarray}

Employing the dispersion relation on both QCD and phenomenological sides, i.e.
\begin{eqnarray}
\Pi(Q^2) &\!\!\!\!=\!\!\!\!& \frac{1}{\pi} \int_0^\infty ds \frac{\text{Im} \Pi(s)}{s + Q^2} + \bigg(\Pi(0) - Q^2 \Pi^\prime(0) \nonumber \\
&\!\!\!\!+\!\!\!\!& \frac{1}{2} Q^4 \Pi^{\prime \prime}(0) - \frac{1}{6} Q^6 \Pi^{\prime \prime \prime}(0) \bigg) \; ,
\end{eqnarray}
where $\Pi(0)$, $\Pi^\prime(0)$, $\Pi^{\prime \prime}(0)$ and $\Pi^{\prime \prime \prime}(0)$ are constants relevant to the correlation function at the origin, then one can establish connection between QCD calculation (the QCD side) and the glueball properties (the phenomenological side),
\begin{eqnarray}
\frac{1}{\pi} \int_0^{\infty} \frac{\mathrm{Im\Pi^{QCD}}(s)}{s + Q^2} ds 
= \frac{f_G^2 M_{0^{--}}^{12}}{M_{0^{--}}^2 + Q^2}  + \int_{s_0}^{\infty} \frac{\rho(s)\theta(s - s_0)}{s + Q^2} ds \ .~~ \label{connection}
\end{eqnarray}

In order to take control of the contributions from higher order condensates in the OPE and the contributions from higher excited and continuum states on phenomenological side, an effective and prevailing way is to perform the Borel transformation simultaneously on both sides of the QCDSR. That is:
\begin{equation}
\hat{B}_{\tau}\equiv \lim_{Q^2\rightarrow \infty,n\rightarrow \infty
\atop
{Q^2\over n}=
{1\over\tau}}\frac{(-Q^2)^n}{(n-1)!}\left(\frac{d}{dQ^2}\right)^n\ ,
\end{equation}
where a parameter $\tau$, usually called the Borel parameter, is introduced.
After performing the Borel transformation, Eq.(\ref{connection}) then turns into
\begin{eqnarray}
\frac{1}{\pi} \int_0^{\infty}\! e^{-s\tau} \mathrm{Im\Pi^{QCD}}(s) ds 
= f_G^2 M_{0^{--}}^{12} e^{- \tau M_{0^{--}}^2} + \! \int_{s_0}^{\infty} \! \rho(s) e^{- s \tau} ds\ .~~
\end{eqnarray}

Taking the quark-hadron duality approximation
\begin{equation}
\frac{1}{\pi}\int_{s_0}^{\infty}
e^{-s\tau}\mathrm{Im\Pi^{QCD}}(s)ds \simeq
\int_{s_0}^{\infty} \rho(s) e^{-s\tau} ds\; ,
\end{equation}
the moments $L_0$ and $L_1$ are achieved,
\begin{eqnarray}
L_0(\tau, s_0) & = & \frac{1}{\pi} \int_0^{s_0} e^{-s\tau} \mathrm{Im\Pi^{QCD}}(s) ds \; , \label{R0} \\
L_1(\tau, s_0) & = & \frac{1}{\pi} \int_0^{s_0} s e^{-s\tau} \mathrm{Im\Pi^{QCD}}(s) ds \; , \label{R1}
\end{eqnarray}
where $L_1$ is obtained via $L_1(\tau, s_0)= - \partial L_0(\tau, s_0)/ \partial \tau$.
Then the $0^{--}$ oddball mass is obtained in form of the ratio of  $L_1(\tau, s_0)$ to $L_0(\tau, s_0)$, i.e.
\begin{eqnarray}
M^i_{0^{--}}(\tau, s_0) = \sqrt{ \frac{L_{1}(\tau, s_0)}{L_{0}(\tau, s_0)}} \label{mass}
\end{eqnarray}
with $i$ for cases $A, B, C$ and $D$.

To evaluate the oddball mass numerically, the following inputs are adopted \cite{Hao:2005hu}:
\begin{eqnarray}
\begin{aligned}
& \langle \alpha_s G^2\rangle = 0.06 \, \text{GeV}^4 \; ,\;
\langle g_s G^3\rangle = (0.27 \, \text{GeV}^2) \langle \alpha_s G^2\rangle \; ,\\
& ~~~~~~~~~~~\Lambda_{\overline{\text{MS}}} = 300 \, \text{MeV} \; ,\;
\alpha_s = \frac{-4\pi}{11 \ln (\tau \Lambda^2_{\overline{\text{MS}}})} \; ,
\end{aligned}
\end{eqnarray}
where the magnitude of trigluon condensate, $\langle g_s G^3 \rangle$, is obtained from the dilute gas instanton model due to the lack of direct knowledge from experiment, while other parameters are commonly used in the literature.

\begin{figure}[hbpt]
\begin{center}
\includegraphics[width=6.1cm]{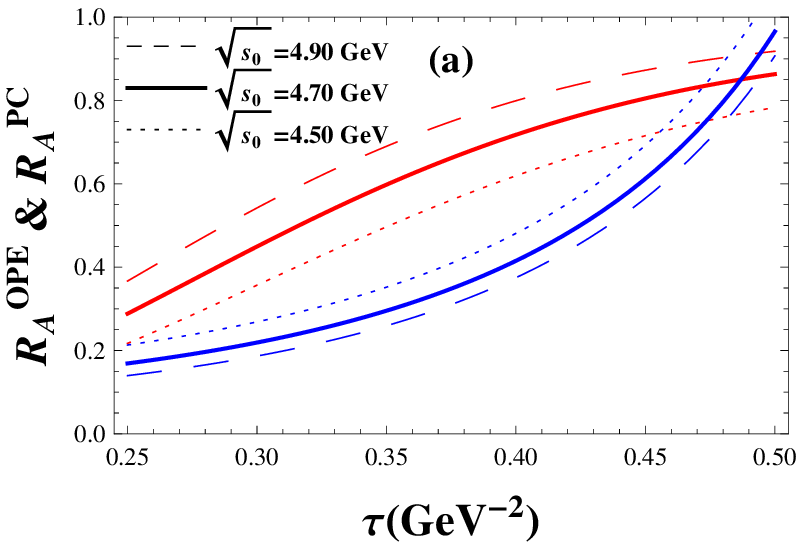}
\includegraphics[width=6.1cm]{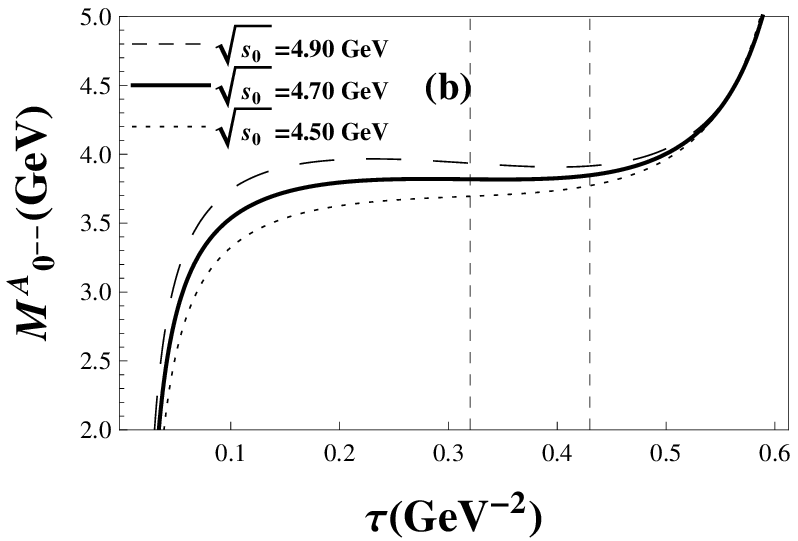}
\caption{(a) The ratios $R^{OPE}_A$ and $R^{PC}_A$ in case-A as functions of Borel parameter $\tau$ for different values of $\sqrt{s_0}$, where blue lines represent $R^{OPE}_A$ and red lines denote $R^{PC}_A$. (b) The mass $M_{0^{--}}^A$ as function of the Borel parameter $\tau$ for different values of $\sqrt{s_0}$, where the two vertical lines indicate the upper and lower limits of the valid Borel window. }
\label{fig-1}
\end{center}
\end{figure}

\begin{figure}[hbpt]
\begin{center}
\includegraphics[width=6.1cm]{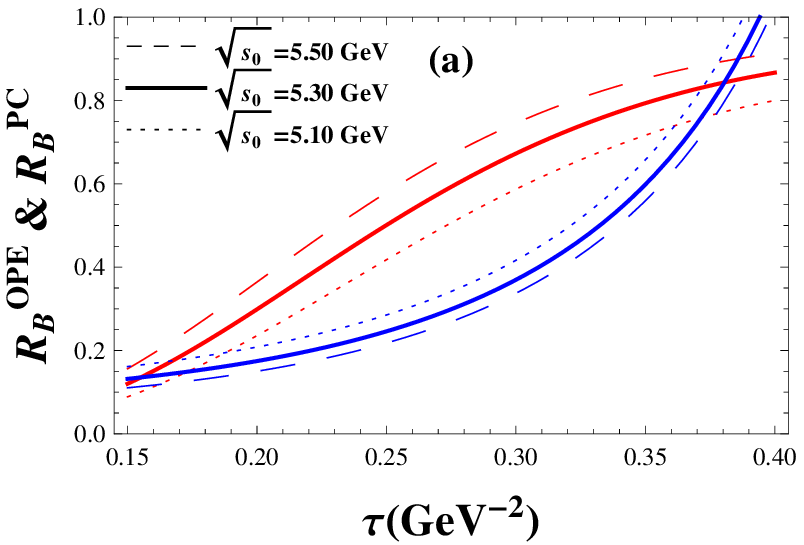}
\includegraphics[width=6.1cm]{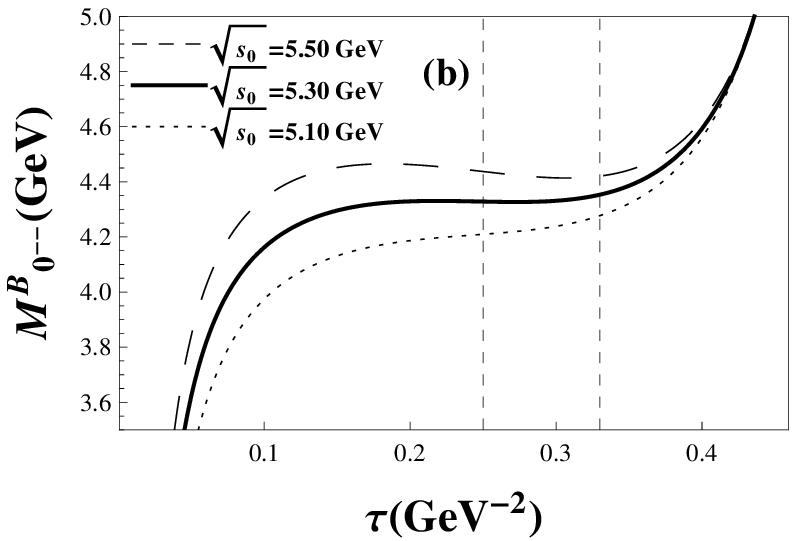}
\caption{The same caption as in Figure 1, but for case-B.}
\label{fig-2}
\end{center}
\end{figure}

\begin{figure}[hbpt]
\begin{center}
\includegraphics[width=6.1cm]{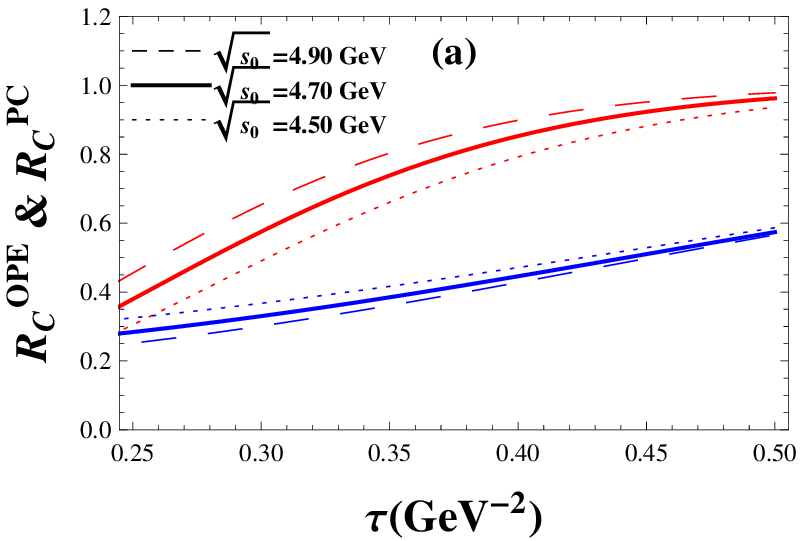}
\includegraphics[width=6.1cm]{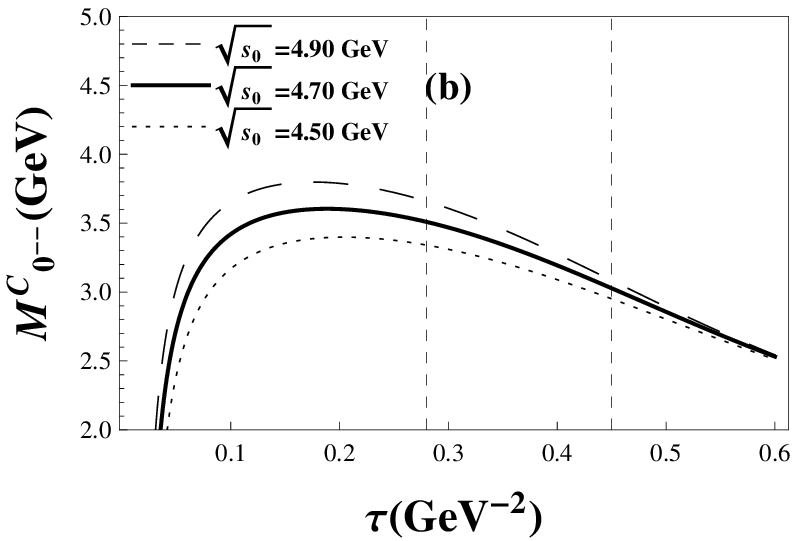}
\caption{The same caption as in Figure 1, but for case-C.}
\label{fig-3}
\end{center}
\end{figure}

\begin{figure}[hbpt]
\begin{center}
\includegraphics[width=6.1cm]{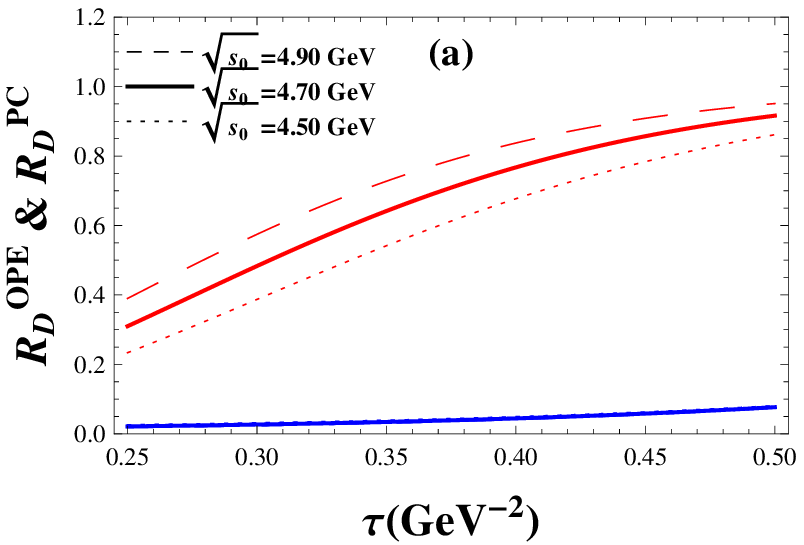}
\includegraphics[width=6.1cm]{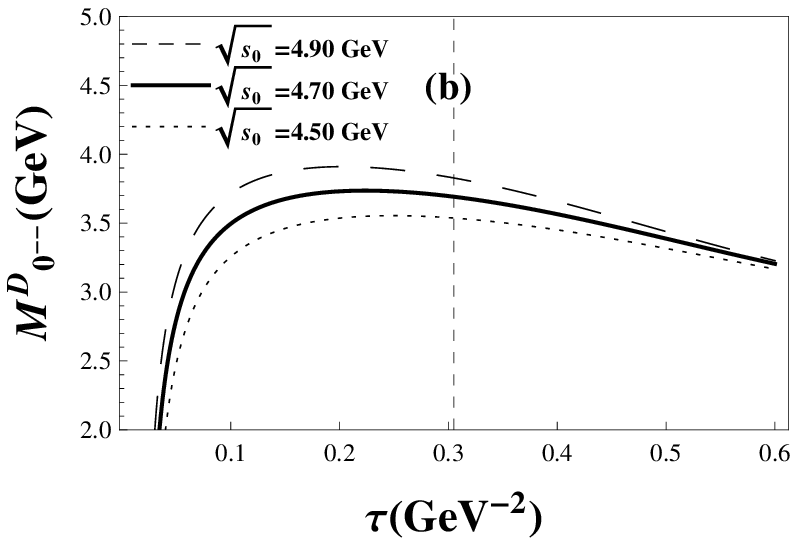}
\caption{The same caption as in Figure 1, but for case-D. Here the single vertical line indicates the lower limit of the valid Borel window while the upper limit is out of the region.}
\label{fig-4}
\end{center}
\end{figure}

\begin{table}[hbpt]
\caption{The lower and upper limits of the Borel parameter $\tau$ (GeV$^{-2}$) for $0^{--}$ oddballs for various cases with different $\sqrt{s_0}$ (GeV).}
\begin{center}
\renewcommand\arraystretch{1.1}
\begin{tabular}{|c|c|c||c|c|c||c|c|c||c|c|c|}
\hline \multicolumn{3}{|c||}  {case-A}  &  \multicolumn{3}{|c||}
{case-B} & \multicolumn{3}{|c||}  {case-C} & \multicolumn{3}{|c|}  {case-D}\\
\hline $\sqrt{s_0}$ & $\tau_{min}$  & $\tau_{max}$ & $\sqrt{s_0}$  & $\tau_{min}$ & $\tau_{max}$ & $\sqrt{s_0}$ & $\tau_{min}$  & $\tau_{max}$ & $\sqrt{s_0}$  & $\tau_{min}$ & $\tau_{max}$\\
\hline 4.90 & 0.29 & 0.44 & 5.50 & 0.23 & 0.34 & 4.90 & 0.26 & 0.45 & 4.90 & 0.28 & 0.86\\
\hline 4.70 & 0.32 & 0.43 & 5.30 & 0.25 & 0.33 & 4.70 & 0.28 & 0.45 & 4.70 & 0.31 & 0.86\\
\hline 4.50 & 0.36 & 0.41 & 5.10 & 0.27 & 0.32 & 4.50 & 0.30 & 0.44 & 4.50 & 0.34 & 0.86\\
\hline
\end{tabular}
\end{center}
\label{window}
\end{table}

In the QCDSR calculation, the parameter $\tau$ and the threshold $s_0$ are free parameters, proceeding from some requirements. Conventionally, two criteria are adopted in determining the $\tau$ \cite{Shifman, Reinders:1984sr, P.Col, Matheus:2006xi}.  First, the convergence of the OPE should be retained, that is the disregarded power corrections must be small. For this aim, one needs to evaluate the relative weight of each term to the total on the OPE side. Secondly, the pole contribution (PC) should exceed what from the higher excited and continuum states. Therefore, one needs to evaluate the relative pole contribution over total, the pole plus higher excited and continuum states ($s_0 \to \infty$), for various $\tau$. In order to properly eliminate the contribution from higher excited and continuum states, the pole contribution is generally required to be more than $50\%$. The two criteria can be formulated as
\begin{eqnarray}
R^{\text{OPE}}_i = \frac{\int_0^{s_0} e^{-s\tau} \mathrm{Im\Pi^{\langle g_s G^3\rangle}}(s) ds}{\int_0^{s_0} e^{-s\tau} \mathrm{Im\Pi^{QCD}}(s) ds} \label{ratio-OPE}
\end{eqnarray}
and
\begin{eqnarray}
R^{\text{PC}}_i = \frac{L_0(\tau,s_0)}{L_0(\tau,\infty)} \; . \label{ratio-PC}
\end{eqnarray}
Here, $i$ stands for cases $A, B, C$ and $D$, and $\mathrm{Im\Pi^{\langle g_s G^3\rangle}}(s)$ is the imaginary part of the contribution from $\langle g_s G^3 \rangle$. Note that the numerator in $R^{\text{OPE}}_i$ depends only on $\mathrm{Im\Pi^{\langle g_s G^3\rangle}}(s)$; the $\langle \alpha_s G^2 \rangle$ and $\langle \alpha_s G^2 \rangle^2$ give no contribution.

To determine the characteristic value of $\sqrt{s_0}$, we carry out a similar analysis as in Refs.\cite{P.Col, Matheus:2006xi}. Therein, one needs to find out the proper value which has an optimal window for the mass curve of the interested hadron. Within this window, the physical quantity, i.e. the mass of $0^{--}$ oddball, is independent of the Borel parameter $\tau$ as much as possible. Through the above procedure one then obtains the central value of $\sqrt{s_0}$. However, in practice, it is normally acceptable to vary the $\sqrt{s_0}$ by about $0.2 \, \text{GeV}$ in the calculation of the QCDSR, which gives the lower and upper bounds and hence the uncertainties of $\sqrt{s_0}$.

With above preparation we numerically evaluate the mass spectrum of $0^{--}$ oddball. For case-A, we show the ratios $R^{OPE}_A$ and $R^{PC}_A$ as functions of Borel parameter $\tau$ in Fig.\ref{fig-1}(a) with different values of $\sqrt{s_0}$, 4.50 GeV, 4.70 GeV, and 4.90 GeV. The dependency relations between oddball mass $M_{0^{--}}^A$ and parameter $\tau$ are given in Fig.\ref{fig-1}(b). Two vertical lines in Fig.\ref{fig-1}(b) indicate the upper and lower limits of the valid Borel window for the central value of $\sqrt{s_0}$, where a smooth section, the so-called stable plateau, in $M_{0^{--}}^A-\tau$ curve exists, suggesting the mass of possible oddball.
A similar situation for case-B is shown in Fig.\ref{fig-2}(a) and Fig.\ref{fig-2}(b), where the threshold parameters $\sqrt{s_0} = 5.10$ GeV, $5.30$ GeV and $5.50$ GeV. The figure also exhibits a stable plateau in $M_{0^{--}}^B-\tau$ curve, which implies another possible oddball.
The situations for case-C and -D are shown in Fig.(\ref{fig-3}) and Fig.(\ref{fig-4}). We find that no matter what value the $\sqrt{s_0}$ takes, no optimal window for stable plateau exists, where $M_{0^{--}}^C$ or $M_{0^{--}}^D$ is nearly independent of the Borel parameter $\tau$. That means the current structures in Eqs.(\ref{current-C}) and (\ref{current-D}) do not support the corresponding oddballs. Note, in Fig.\ref{fig-4}(b) the upper limit of Borel window is not shown, since it exceeds the region of $\tau$-axis. The exact measures of the Borel windows in four cases are given in Table \ref{window} with various values of $\sqrt{s_0}$.

Our calculation shows that there possibly exist two $0^{--}$ oddballs, corresponding to the currents (\ref{current-A}) and (\ref{current-B}) respectively. That is
\begin{eqnarray}
M_{0^{--}}^A = 3.81 \pm 0.12 \, \text{GeV},
\end{eqnarray}
and
\begin{eqnarray}
M_{0^{--}}^B = 4.33 \pm 0.13 \, \text{GeV},
\end{eqnarray}
where, the errors stem from the uncertainties of Borel parameter
$\tau$ and threshold parameter $\sqrt{s_0}$. From Fig.\ref{fig-1}(b) and Fig.\ref{fig-2}(b), it is obvious that $M_{0^{--}}^A$ and $M_{0^{--}}^B$ are are quite stable and insensitive to the variation of $\tau$ and $\sqrt{s_0}$ within the proper windows of $\tau$. This is the main reason why our calculation yields small errors, similar as Refs.\cite{Huang:1998wj, Narison:1996fm} for instance. Hereafter, we refer these two oddballs as $G_{0^{--}}(3810)$ and $G_{0^{--}}(4330)$ in discussion.

The mass difference of these two $0^{--}$ oddballs are originally due to the different orders of the gluon field strength tensor $G$ and the dual field strength tensor $\tilde {G}$ in Eq.(\ref{current-A}) and Eq.(\ref{current-B}). Note, while these two oddballs will not mix with $q \bar{q}$ states, they can in principle mix with hybrids ($q \bar{q} g$) \cite{General:2006ed} and tetraquark states \cite{Xie:2013uha} with the same quantum number and similar mass, though naively the OZI suppression may hinder the mixing in certain degree. As $G_{0^{--}}(3810)$, $G_{0^{--}}(4330)$ and the $0^{--}$ hybrid meson \cite{General:2006ed} are close in mass, at a minimum 3 state mixing possibility should be further analyzed \cite{QiaoTang}.

Note that result for oddball mass in this work is larger than that in flux tube model, where the mass of a $0^{--}$ oddball was predicted to be about 2.79 GeV \cite{Isgur:1984bm}. Whereas, the lattice QCD calculation yielded an even bigger result with large errors,  $5166 \pm 1000$ MeV \cite{Gregory:2012hu}. In this calculation the instanton and topological charge screening effects have not been taken into account, which as Forkel pointed out is important \cite{twogluon0++}, at least in cases like $0^{++}$ and $0^{-+}$ states. In this work, since the obtained results are very stable and the nonpertubative contributions are already quite large, we speculate the instantons contributions might be small. Detailed analysis on this issue is beyond the scope of this Letter and left for future study.

Experimentally, since the present measurement results for glueball are either contradictory or at least non-conclusive, searching for clear evidence of glueball is now still an outstanding unsolved problem. This situation may be changed if measurement on unconventional glueballs makes progress. We suggest the $0^{--}$ oddballs to be the prior ones in future experimental measurement due to the reason mentioned in above. Following we make a brief analysis on the feasibility of finding oddballs $G_{0^{--}}(3810)$ and $G_{0^{--}}(4330)$ in experiment.

Taking the light one, the $G_{0^{--}}(3810)$, as an example, it can be produced in processes $X(3872) \to \gamma + G_{0^{--}}(3810)$, $\Upsilon(1S) \to f_1 (1285) + G_{0^{--}}(3810)$, $\Upsilon(1S) \to \chi _{c_1} + G_{0^{--}}(3810)$, $\chi_{b_1} \to J/\psi + G_{0^{--}}(3810)$, and $\chi_{b_1} \to \omega + G_{0^{--}}(3810)$. All the parent particles in above processes are copiously produced in experiment, and hopefully decay to the oddball with modest rates. To finally ascertain $G_{0^{--}}(3810)$, a straightforward procedure is to reconstruct it from its decay products, though the detailed characters of it need more work.
Relatively, the exclusive processes are more transparent in this aim,
such as, $G_{0^{--}}(3810) \to \gamma + f_1(1285)$, $G_{0^{--}}(3810) \to \gamma + \chi_{c_1}$, and $G_{0^{--}}(3810) \to \omega + f_1(1285)$. These typical oddball production and decay processes are expected to be measurable in experiments, e.g. at the LHCb. Detailed analysis on these oddballs production and decay issues will be given elsewhere.

In summary, based on the interpolating currents with quantum number of $J^{PC} = 0^{--}$, the oddball mass spectrum is calculated in the framework of QCD Sum Rules. Two stable $0^{--}$ oddballs are obtained with masses about $3.81 \, \text{GeV}$ and $4.33 \, \text{GeV}$. We have briefly analysed the $0^{--}$ oddball optimal production and decay mechanisms, which indicates that the long search elusive glueball is expected to be measured in BESIII, BELLEII, Super-B, PANDA, and LHCb experiments.

\acknowledgments

The authors acknowledge for the helpful discussion with C.-Z. Yuan, J.-Z. Zhang and R.-G. Ping. This work was supported in part by the National Natural Science Foundation of China(NSFC) under the grants 11175249, 11121092, and 11375200.

\end{document}